# Evidence Against Correlations Between Nuclear Decay Rates and Earth-Sun Distance

Eric B. Norman[1,2,3], Edgardo Browne[3], Howard A. Shugart[4],
Tenzing H. Joshi[1], and Richard B. Firestone[3]

[1] *Nuclear Engineering Department, University of California, Berkeley, CA 94720*
[2] *N-Division, Lawrence Livermore National Laboratory, Livermore, CA 94551*
[3] *Nuclear Science Division, Lawrence Berkeley National Laboratory, Berkeley, CA 94720*
[4] *Physics Department, University of California, Berkeley, CA 94720*

## Abstract

We have reexamined our previously published data to search for evidence of correlations between the rates for the alpha, beta-minus, beta-plus, and electron-capture decays of $^{22}$Na, $^{44}$Ti, $^{108}$Ag$^m$, $^{121}$Sn$^m$, $^{133}$Ba, and $^{241}$Am and the Earth-Sun distance. We find no evidence for such correlations and set limits on the possible amplitudes of such correlations substantially smaller than those observed in previous experiments.

Corresponding author: Eric B. Norman
E-mail address: ebnorman@lbl.gov

Several careful experiments designed to study the decays of long-lived radioactive isotopes have reported observations of small periodic annual variations modulating the well-known exponential decay curve.[1,2,3] Recently, Jenkins et al.[4] proposed that these decay rate variations were correlated with the distance between the Earth and the Sun. Jenkins et al. went on to suggest that the underlying mechanism responsible for this correlation might be some previously unobserved field emitted by the Sun or perhaps was the result of the annual variation in the flux of solar neutrinos reaching the Earth. If the Jenkins et al.[4] proposal were correct, it would have profound consequences for many areas of science and engineering. Thus, it is important to test this proposal in a variety of experiments. Therefore, we have reanalyzed a large body of decay data that we collected over the past 15 years to search for the type of periodic variations observed in the above-mentioned experiments. The data we examined was collected in three separate gamma-ray experiments that were designed to measure the half-lives of $^{44}$Ti, $^{121}$Sn$^m$, and $^{108}$Ag$^m$ (Refs. 5,6,7).

In the first of these experiments, gamma-ray spectra from a mixed source containing $^{22}$Na and $^{44}$Ti were collected for a period of approximately 2 years to determine the half-life of $^{44}$Ti (Ref. 5). $^{22}$Na decays via beta-plus emission and electron capture whereas $^{44}$Ti decays via electron capture only. A 110-cm$^3$ high-purity germanium detector coupled to an ORTEC ACE data acquisition system was used to collect γ-ray spectra in 4096 channels in 1-day intervals starting on January 14, 1994. Ten consecutive spectra were then summed together for data analysis. A similar experiment was carried out to determine the half-life of $^{121}$Sn$^m$ (Ref. 6). In this case a mixed source of $^{121}$Sn$^m$ and $^{241}$Am was counted using a 36-mm diameter x 13-mm thick planar germanium detector. $^{121}$Sn$^m$ decays via beta-minus emission whereas $^{241}$Am decays via alpha emission. Gamma-ray spectra were collected for 450 days in 4-day intervals beginning July 24, 2000. Finally, an experiment was also conducted to determine the half-life of $^{108}$Ag$^m$ (Ref. 7). For this measurement a mixed source of $^{44}$Ti, $^{108}$Ag$^m$, and $^{133}$Ba was counted for more than one year using the same detector system as used in the $^{121}$Sn$^m$ experiment. $^{108}$Ag$^m$ decays primarily via electron capture and $^{133}$Ba decay exclusively by electron capture. Gamma-ray spectra were collected in one-week intervals beginning January 27, 2003.

To extract the net areas of the gamma-ray peaks of interest from each spectrum, we selected a peak region and then background regions of the same energy width above and below the peak. The average of the two background areas was then subtracted from the peak area to obtain the net peak area. The half-lives that we measured in these experiments are all quite long: ~ 60 years for $^{44}$Ti, ~ 40 years for $^{121}$Sn$^m$, and ~ 400 years for $^{108}$Ag$^m$. Thus even over the lengthy data taking periods, the change in activity of each of these radioisotopes is small (~ 1% or less per year). In order to minimize the influence of any changes in detector and/or electronics performance, we analyzed ratios of gamma-ray peak areas from the isotope of interest and those from a reference isotope whose half life was well known. Analyses of the time dependences of these ratios allowed us to determine the half lives of $^{44}$Ti, $^{121}$Sn$^m$, and $^{108}$Ag$^m$ (Refs. 5,6,7). As a result of similar concerns regarding the long-term stability of their experimental apparatus, this same type of approach was used by Alburger *et al.*[1] in their measurement of the half-life of $^{32}$Si relative to that of $^{36}$Cl. If the Jenkins proposal were correct, it is very unlikely that the alpha, beta-minus, beta-plus, and electron-capture decays of all radioactive isotopes would be affected in quantitatively the same way. Thus the ratios of counts observed from two different isotopes would also be expected to show annual variations.

In order to search for variations in the decay rates of these isotopes that might be correlated with the Earth-Sun distance, $R_{ES}$, we corrected the observed ratios for the expected exponential decays of both isotopes. In practice, this meant multiplying each measured ratio by $\exp(+\lambda_{eff} t)$, where $\lambda_{eff} = \lambda_1 - \lambda_2$. The results of this procedure for the data obtained in our three experiments are illustrated in

Figures 1, 2, and 3. In each case, the data points have been normalized to the mean value of each data set, and the horizontal line at 1.00 represents this mean value. Thus, the horizontal line is our model for the null hypothesis (i.e., no annual variation). The amplitude of the periodic variations observed by Alburger et al.[1] in the ratio of $^{32}$Si/$^{36}$Cl beta-minus counting rates and by Siegert et al.[2] in the alpha counting rate of $^{226}$Ra were both approximately 0.15%, whereas that reported by Falkenberg[3] was 0.37%. The oscillatory curve shown in each figure represents the variation in $1/R^2_{ES}$ over the experimental data acquisition period reduced by a factor of 20. The amplitude of these annual variations is thus 0.15%. We identify this curve as the "Jenkins hypothesis".

From the $^{22}$Na/$^{44}$Ti data set, we calculate $X^2_\nu = 1.08$ for 63 degrees of freedom for the null hypothesis (i.e. no annual variation); $X^2_\nu = 3.39$ for the Jenkins hypothesis. From the $^{241}$Am/$^{121}$Sn$^m$ data set we find $X^2_\nu = 1.09$ for 69 degrees of freedom for the null hypothesis; $X^2_\nu = 5.25$ for the Jenkins hypothesis. From the $^{133}$Ba/$^{108}$Ag$^m$ data set we find $X^2_\nu = 1.23$ for 58 degrees of freedom for the null hypothesis; $X^2_\nu = 20.8$ for the Jenkins hypothesis. In all of the cases we have studied, the null hypothesis is strongly favored over the Jenkins hypothesis. By examining the change in the total $X^2$ for each data set as the amplitude of the annual variations is varied from 0, we set $3\sigma$ upper limits (i.e., $\Delta(X^2) = 9$) of 0.06%, 0.024%, and 0.004%, respectively, on the amplitude of an annual variation in these decays rates that is correlated with the Earth-Sun distance.

It is interesting to note that in the work of Alburger et al.[1] a very statistically significant annual variation in the ratio of count rates of $^{32}$Si/$^{36}$Cl was observed (see Fig. 4 of Ref. 1). These authors could not identify a mechanism that could quantitatively explain these observations. However, in the work of Siegert et al.[2], where a similar amplitude annual variation in the count rate of $^{226}$Ra was observed (see Fig. 1 of Ref. 2), when the ratio of observed count rates of $^{154}$Eu/$^{226}$Ra was examined, the annual variations disappeared (see Fig. 3 of Ref. 2). These authors attributed the annual variations observed from the decays of a single source to a yearly variation in the performance of their experimental equipment that cancelled out in the ratio.

In conclusion, we find no evidence for correlations between the rates for the decays of $^{22}$Na, $^{44}$Ti, $^{108}$Ag$^m$, $^{121}$Sn$^m$, $^{133}$Ba, and $^{241}$Am and the Earth-Sun distance. We set limits on the possible amplitudes of such correlations (2.5 – 37) times smaller than those observed in previous experiments[1,2,3]. Our results strongly disfavor the suggestions by Jenkins *et al.*[4] of an annual variation based on a previously unobserved field produced by the Sun or the annual variation in the flux of solar neutrinos reaching the Earth. Very recently, Cooper[8] performed a very clever analysis of decay power data obtained from the $^{238}$Pu thermoelectric generator aboard the Cassini spacecraft. The results of this analysis also strongly disagree with the hypothesis of a correlation between nuclear decay rates and the distance of the source to the Sun.

This work was supported in part by the U. S. Department of Homeland Security, and by the U. S. Department of Energy under contract numbers DE-AC52-07NA27344 at LLNL and DE-AC02-05CH11231 at LBNL.

# Figure Captions

1. Data points represent the normalized ratio between the 1274- and 1157-keV gamma-ray peak areas from $^{22}$Na and $^{44}$Ti reported in Ref. 5 corrected for the exponential decays of both isotopes. The horizontal line at a value of 1.00 represents the mean of all the data points. The oscillatory curve (Jenkins hypothesis) represents the variation in $1/R^2_{ES}$ over the experimental data acquisition period divided by a factor of 20. $X^2_\nu = 1.08$ for 63 degrees of freedom for the null hypothesis (i.e. no annual variation); $X^2_\nu = 3.39$ for the Jenkins hypothesis.

2. Data points represent the normalized ratio between the 59- and 37-kev gamma-ray peak areas from $^{241}$Am and $^{121}$Sn$^m$ reported in Ref. 6 corrected for the exponential decays of both isotopes. The horizontal line at a value of 1.00 represents the mean of all the data points. The oscillatory curve (Jenkins hypothesis) represents the variation in $1/R^2_{ES}$ over the experimental data acquisition period divided by a factor of 20. $X^2_\nu = 1.09$ for 69 degrees of freedom for the null hypothesis (i.e. no annual variation); $X^2_\nu = 5.25$ for the Jenkins hypothesis.

3. Data points represent the normalized ratio between the 356- and the efficiency-weighted sum of the 434-, 614- and 723-kev gamma-ray peak areas from $^{133}$Ba and $^{108}$Ag$^m$ reported in Ref. 7 corrected for the exponential decays of both isotopes. The horizontal line at a value of 1.00 represents the mean of all the data points. The oscillatory curve (Jenkins hypothesis) represents the variation in $1/R^2_{ES}$ over the experimental data acquisition period divided by a factor of 20. $X^2_\nu = 1.23$ for 58 degrees of freedom for the null hypothesis (i.e. no annual variation); $X^2_\nu = 20.8$ for the Jenkins hypothesis.

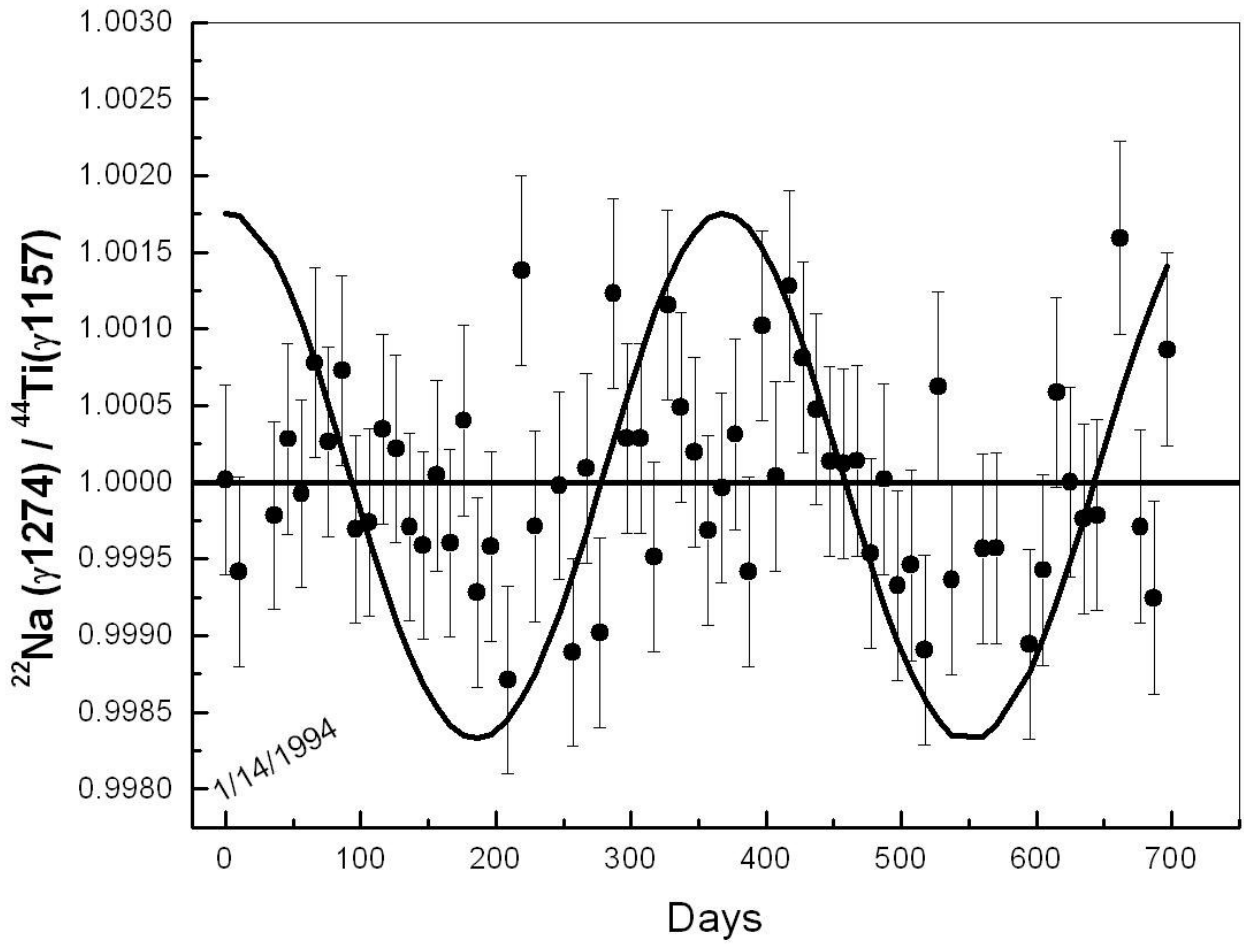

Figure 1

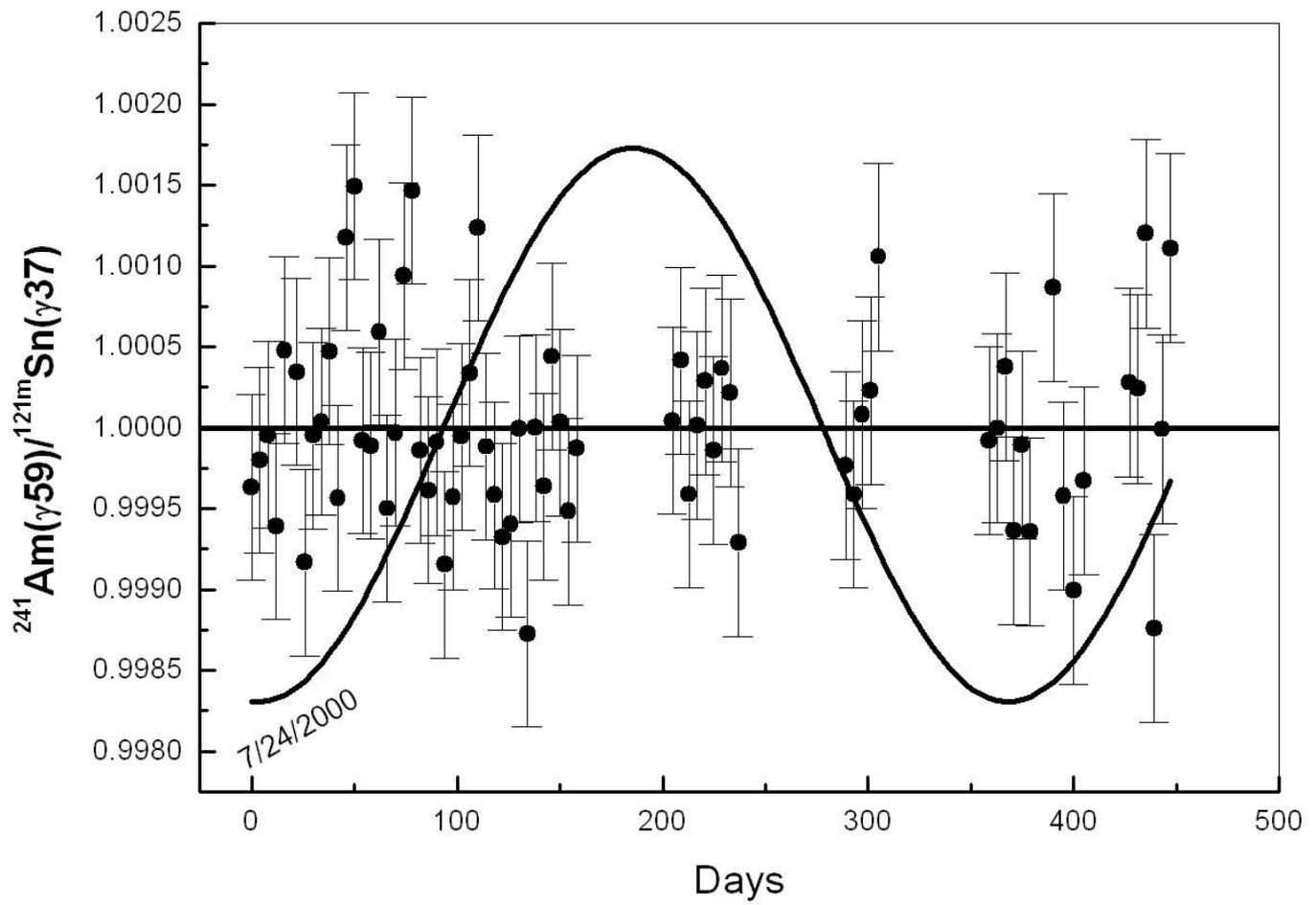

Figure 2

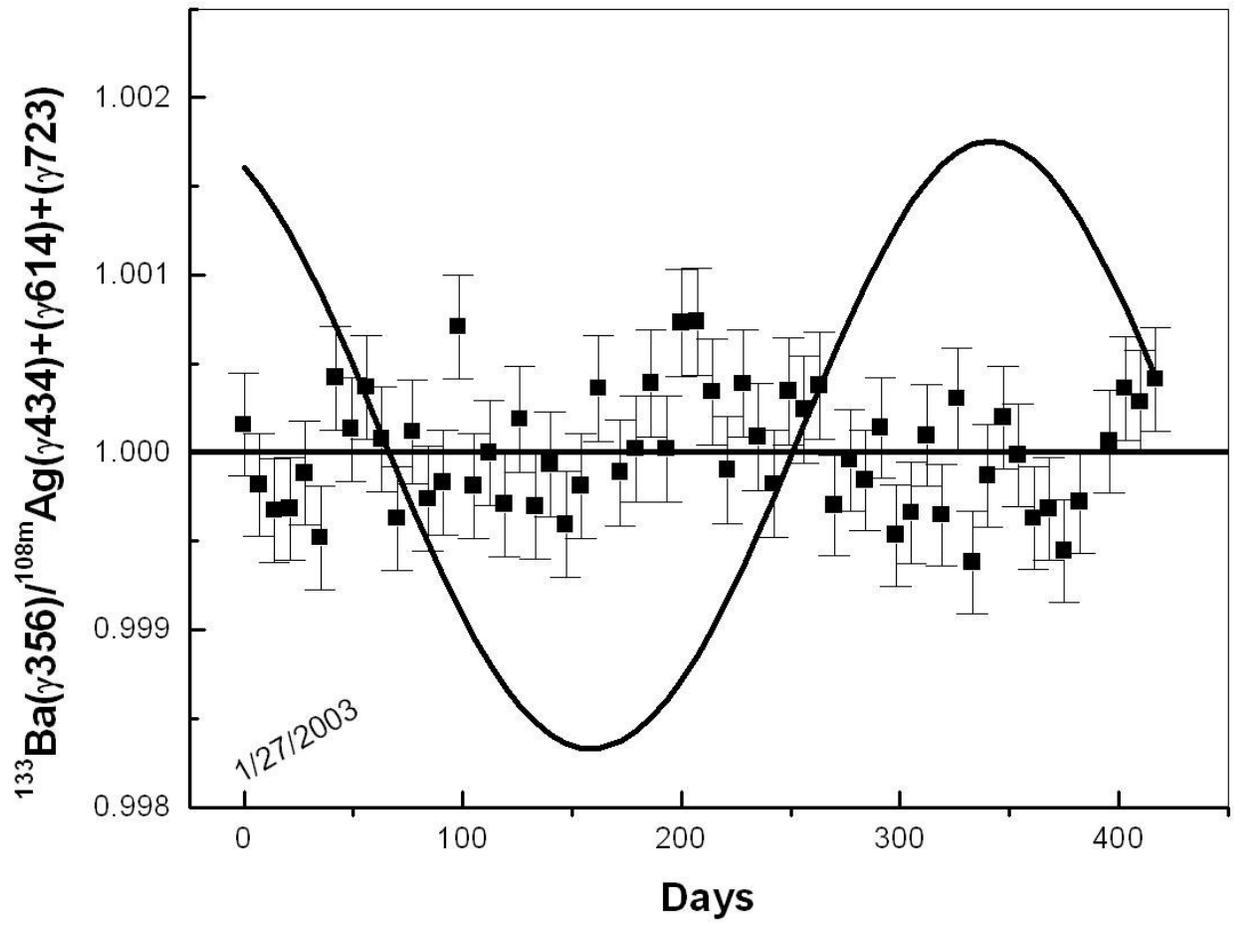

Figure 3